\RequirePackage{fix-cm}
\documentclass[twocolumn]{svjour3}          % twocolumn
\smartqed  % flush right qed marks, e.g. at end of proof
\usepackage{graphicx}
\usepackage{bm}
\usepackage{psfrag}
\begin{document}

\title{Dynamics of an Intruder in Dense Granular Fluids
\thanks{dedicated to Isaac Goldhirsch}
}
%\subtitle{A study of the dynamics caused by a directed force on a particle as well as (of) the dependence on momentum propagation in a granular medium}

\titlerunning{Dynamics of an Intruder in a Dense Granular Medium}        
% if too long for running head

\author{Andrea Fiege         \and
        Matthias Grob \and
        Annette Zippelius
}


\institute{Andrea Fiege \at
              %Max-Planck-Institut f\"ur Dynamik und Selbstorganisation, Bunsenstr. 10, D-37073 G\"ottingen, Germany \\
              Institut f\"ur Theoretische Physik, Friedrich-Hund-Platz 1, D-37077 G\"ottingen, Germany \\
	      Max-Planck-Institut f\"ur Dynamik und Selbstorganisation, Bunsenstr. 10, D-37073 G\"ottingen, Germany 
	      \email{fiege@theorie.physik.uni-goettingen.de}
\and
Matthias Grob \at
              Institut f\"ur Theoretische Physik, Friedrich-Hund-Platz 1, D-37077 G\"ottingen, Germany 
              %\email{@example.com}           %  \\
\and
	      Annette Zippelius \at
              Institut f\"ur Theoretische Physik, Friedrich-Hund-Platz 1, D-37077 G\"ottingen, Germany \\
	      Max-Planck-Institut f\"ur Dynamik und Selbstorganisation, Bunsenstr. 10, D-37073 G\"ottingen, Germany 
}

\date{Received: date / Accepted: date}

\maketitle

\begin{abstract}


We investigate the dynamics of an intruder pulled by a constant force
in a dense two-dimensional granular fluid by means of event-driven
molecular dynamics simulations. In a first step, we show how a
propagating momentum front develops and compactifies the system when
reflected by the boundaries. To be closer to recent experiments
\cite{candelier2010journey,candelier2009creep}, we then add a frictional
force acting on each particle, proportional to the particle's
velocity. We show how to implement frictional motion in an event-driven
simulation. This allows us to carry out extensive numerical
simulations aiming at the dependence of the intruder's velocity on
packing fraction and pulling force. We identify a linear relation for
small and a nonlinear regime for high pulling forces and investigate
the dependence of these regimes on granular temperature.
\end{abstract}

\section{Introduction}
\label{sec:intro}

The response of an intruder to a pulling force is a rather versatile
tool to study the local nonequilibrium dynamics in various complex
fluids,
such as glasses \cite{hastings2003depinning}, colloids \cite{habdas04,gazuz09} and 
granular media \cite{candelier2010journey,candelier2009creep,zik92,sarracino2010irreversible}. The measured 
force-velocity relations reveal striking nonlinear behaviour close to
the glass and/or jamming transition. In the early experiments on
colloids evidence was given for a threshold force even in the fluid regime,
whereas mode-coupling theory predicts such
behaviour only in the glassy phase.
In the experiment of
Ref. \cite{candelier2010journey,candelier2009creep} two transitions
are observed: the first one, called fluidization, separates a regime
of continuous to intermittent motion of the intruder. It occurs below
the jamming point (the second transition) and depends on the applied
pulling force. Experimentally it is observed even for the smallest
pulling force with the possibility of a dynamic transition inherent to
the vibrated granular fluid for zero applied force.
In Ref. \cite{reichhardt2010}, the dynamics of an intruder was
simulated near the jamming point. One result of this study are
velocity-force relations which are linear for packing fraction, 
$\eta\leq 0.833$ and become nonlinear for $\eta$ still closer to the
jamming point.

In contrast to \cite{reichhardt2010}, we discuss a stochastically
driven system, describing a fluidized granular state -- similar to the
experimental setup in \cite{candelier2010journey,candelier2009creep}.
A recent mode-coupling theory \cite{kranz2010} predicts that such a
``thermalized'' granular fluid undergoes a glass transition at a
packing fraction below the jamming point. This transition is different
from both, the jamming transition at zero temperature and the glass
transition for either Newtonian or Brownian dynamics.

In this paper we
use event driven simulations to analyze the dynamics of an intruder in
a two-dimensional system of hard disks close to the glass transition. We compute
force-velocity correlations in the linear and nonlinear regime,
extract the mobility for the linear regime and discuss scaling for the nonlinear regime. Moreover we discuss the dependence on the granular temperature.

%Such a ``thermalized'' granular fluid may undergo a glass transition,
%presumably at a filling fraction below the jamming point
%\cite{kranz2010}. 

\section{Model}
\label{sec:model}
We consider a bidisperse system of $N$ hard disks in an area
$A$. Particles' positions and velocities are denoted by $\{\mathbf
r_i\}_{i=1}^N$ and $\{\mathbf v_i\}_{i=1}^N$. The size ratio
$R_s/R_b=4:5$ of small to big particles is chosen as in the
experiments of Refs. \cite{candelier2010journey,candelier2009creep}. The mass ratio
follows from the disk like shape of the particles as $m_s/m_b=16/25$.
The intruder with position $\mathbf r_0$ and velocity $\mathbf v_0$ is chosen
twice as large as the small particles, $R_0=2R_s$ and $m_0=4m_s$.

%\paragraph{Inelastic collsions}
The particles collide inelastically so that in each collision energy is
dissipated while momentum is conserved. The simplest model ignores 
rotational degrees of freedom, allowing for normal restitution only.
The collision rules for two colliding particles (say $_1$ and $_2$)
is given in terms of their relative velocity $\mathbf g:=\mathbf
v_1-\mathbf v_2$
\begin{equation}
  \label{eq:coll1}
  (\mathbf g \cdot \mathbf n)' = -\epsilon (\mathbf g \cdot \mathbf n).
\end{equation}
where $\epsilon$ is the coefficient of restitution and $\mathbf
n$ denotes the unit vector $\mathbf n:=(\mathbf r_1-\mathbf
r_2)/|\mathbf r_1-\mathbf r_2|$ at contact.

In between collisions the particles perform Brownian motion, due to
friction with a surrounding medium or with the bottom plate. We model
this by a frictional force in the equation of motion: $\mathbf
F_{\mathrm{fr}}=-\gamma m \mathbf v$. For all finite values of the friction
coefficient $\gamma$, momentum is lost, whereas for $\gamma=0$,
the total momentum is conserved. 

%\paragraph{Driving}
Inelastic collisions give rise to a loss of energy $\propto \frac{1 -
  \epsilon^2}{2}T$, where $T$ denotes the granular temperature
$T=\frac{1}{2N}\sum_{i=1}^{N}m_i v_i^2$. To balance this dissipation the entire system is
driven stochastically (like an air-fluidized bed \cite{oger-1996,abate-2006,abate-07} or a vibrating bottom
plate \cite{urbach-2005,reis-2006,reis-2007}) by instantaneous kicks. The momentum of the $i$-th particle
$\mathbf p_i(t)$ is changed according to
\begin{equation}
  \label{eq:drive1}
\mathbf  p_i'(t) =\mathbf  p_i(t) + p_{\mathrm{dr}} \bm\xi_i(t)
\end{equation}
with $\langle \xi_i^{\alpha}(t)\xi_j^{\beta}(t') \rangle =
\delta_{ij}\delta_{\alpha \beta}\delta (t - t')$ and zero mean
$\langle \xi_i^{\alpha}(t) \rangle = 0$, and $\alpha=x,z$ denoting the cartesian components.
In the frictional case, $\gamma\neq 0$, we kick single particles,
whereas for the frictionless case, $\gamma= 0$, we kick neighbouring
particles with equal and opposite momenta in order to conserve the
total momentum locally \cite{fiege-2009}.

% In order to balance
%the loss due to collisions and the gain due to stochastic driving the
%driving frequency is chosen to be equal the \emph{Enskog frequency}
%$\nu_{dr} \approx \omega_E$. 

%\paragraph{Pulling force}
We want to investigate the dynamics of a tagged particle under the
action of a deterministic pulling force $\mathbf F$. This so called
{\it intruder} with position $\mathbf r_0$ and velocity $\mathbf v_0$
is subject to systematic kicks
 \begin{equation}
  \label{eq:drive2}
 \mathbf p_0'(t) = \mathbf p_0(t) + \mathbf F \Delta t
\end{equation}
mimicking a constant force acting on the intruder. The frequency of these systematic kicks $(\Delta t)^{-1}$ is chosen
4 orders of magnitude higher than the frequency of the stochastic driving and the collision frequency (see below). The systematic force does work on the system, injecting momentum and energy.

Combining inelastic collisions, stochastic driving, pulling force and friction, we arrive at the following equation of motion
\begin{equation}
  \label{eq:com2}
  \dot{p}_i^{\alpha}(t) = - \gamma p_i^{\alpha}(t) +
  F\delta_{\alpha,x}\delta_{i,0} +  \left. \frac{\mathrm{d} p_i^{\alpha}}{\mathrm{d}t}  \right |_{\mathrm{dr}}
  + \left. \frac{\mathrm{d} p_i^{\alpha}}{\mathrm{d}t}  \right |_{\mathrm{coll}}.
\end{equation}

%\paragraph{Stationary states} 
%\label{sec:stationarity}

We aim to prepare the granular fluid without pulling force to be under
stationary conditions. This can be achieved by balancing energy
dissipation due to friction and inelastic collisions by the
stochastic driving. Without external forcing the mean velocity of the
particles vanishes, so that the granular temperature,
$T=\frac{1}{2N}\sum_{i=1}^{N}m_i v_i^2$, is just the average
kinetic energy of the particles. Its time rate of change can be deduced from Eq. \ref{eq:com2} by multiplying with ${p}_i^{\alpha}(t)$ yielding
\begin{equation}
 \frac{1}{2} \frac{d}{dt}\left({p}_i^{\alpha}\right)^2(t) = - \gamma (p_i^{\alpha}(t))^2 + \frac{1}{2}  \left. \frac{\mathrm{d} }{\mathrm{d}t}  \right |_{\mathrm{dr}}(p_i^{\alpha})^2
  + \frac{1}{2}\left. \frac{\mathrm{d} }{\mathrm{d}t}  \right |_{\mathrm{coll}}(p_i^{\alpha})^2.
\end{equation}
The explicit calculation of the time derivations on the rhs are beyond the scope of this paper but may be found elsewhere \cite{aspelmeier2001free,kranz2011}. The time rate of change of the granular temperature then reads as follows:
\begin{equation} 
\frac{\mathrm{d} T}{\mathrm{d}t} = -2\gamma T -\omega_E\frac{1-\epsilon^2}{2}T+ 
f_{\mathrm{dr}}\frac{p_{\mathrm{dr}}^2}{2m_{\mathrm{eff}}},
\end{equation}
where the effective mass is given by
$m_{\mathrm{eff}}=\frac{m_sm_b}{m_s+m_b}$. For simplicity, we choose the
driving frequency $f_{\mathrm{dr}}$ equal to the Enskog frequency $\omega_E$ \cite{boon1991molecular} and
measure length and mass in units, such that $R_s=1$ and $m_s=1$.
In the stationary state, $\frac{\mathrm{d} T}{\mathrm{d}t} =0$, the
amplitude of the stochastic driving is given by:
\begin{equation}
\frac{p_{\mathrm{dr}}^2}{2m_{\mathrm{eff}}}=\frac{2\gamma T}{\omega_E}+\frac{1-\epsilon^2}{2}T.
\end{equation}
We prepare our system in a stationary state with $T=1$, yielding a numerical value for $p_{\mathrm{dr}}$ depending on $\gamma$, $\omega_E$ and $\epsilon$.

\section{Simulations}

\subsection{Implementation as event-driven simulation}

In order to apply an event-driven algorithm to the dynamics, as
described by Eq. (\ref{eq:com2}), we need to generalize the code to account for
damping ($\gamma\neq 0$). As usual, events include collisions of
particles, wall collisions, subbox-wall collisions and (discrete)
driving events (kicks). Standard event driven algorithms calculate the
time of an upcoming event and advance all particles to that time under
the assumption that the particle motion is ballistic. Hence, in
between events, the velocities do not change. The condition for a
collision is $R_i + R_j = \left\vert \mathbf r_i(t^\bullet)- \mathbf
  r_j(t^\bullet) \right\vert $, i.e., the difference in the particles'
trajectories at the collision time $t^\bullet$ must be equal to the
sum of their radii. Plugging in the trajectories subject to ballistic
motion $\mathbf r_{i,j}(t^\bullet) = \mathbf r_{i,j}(t_0) + \mathbf
v_{i,j}(t_0)(t^\bullet-t_0)$, one finds a quadratic equation, that
must have a positive solution for $(t^\bullet-t_0)$ if and only if a collision
will occur \cite{alder-1959,Lubachevsky-1991}.

In our case, the velocities decrease as $\mathbf v_{i,j} (t^\star) =
\mathbf v_{i,j}(t_0) \exp(-\gamma (t^\star-t_0))$. Integration results
in
\begin{equation}
  \mathbf r_{i,j}(t^\star) =  \mathbf r_{i,j} (t_{0}) +  
\mathbf v_{i,j}(t_{0})\frac{1 - e^{- \gamma(t^\star-t_{0})}}{\gamma}.
\end{equation}
Inserting this into the condition for a collision, we find a similar
condition for a collision except that $(t^\bullet-t_0)$ is replaced by
$\frac{1 - e^{- \gamma(t^\star-t_{0})}}{\gamma}$, which is monotonically 
increasing in $(t^\star-t_0)$. Since our
(ballistic) event-driven code calculates $(t^\bullet-t_0)$ we only
have to replace the collision time by
\begin{equation}
t^\star -t_0 = -\frac{1}{\gamma} 
\log\left( 1- \gamma \cdot(t^\bullet-t_0) \right). \label{eq:colltime}
\end{equation}
Hence, collisions will occur at the same place and in the same order but at a different
time and with different particles' velocities, provided the place of the collision is within reach of the particle. In detail, the maximum distance a particle is able to travel is limited because of the friction slowing the particle down. The maximal distance  is given by $\mathbf r_i
^{max}=\lim_{t^\star \rightarrow \infty} \mathbf r_i(t^\star)=
\mathbf r_i(t_0) + \mathbf v_i(t_0)/\gamma > \mathbf r_i(t^\bullet) $
that must be larger than the distance to the place at which the event
will take place. Hence, $(\mathbf r_i(t^\bullet)-\mathbf r_i
(t_0))<\mathbf v_i(t_0)/\gamma$ or $(t^\bullet -t_0)<1/\gamma$. This
is consistent with Eq. (\ref{eq:colltime}) which requires\\ 
$\log\left(
  1- \gamma \cdot(t^\bullet-t_0) \right)$ to be negative for a
positive collision time. Advancing a particle to a different type of event ( e.g. kicks) is done in the same way.

Changing an existing event-driven simulation in the way discussed
enables us to simulate systems of hard disks and spheres subject to
friction almost as fast as without friction. The systematic force on
the intruder as described in Eq. \ref{eq:drive2} is implemented as
frequent kicks on the intruder. In order to avoid an inelastic collapse, i.e., a infinite number of collisions
in a finite time interval, we use the same method as described in
\cite{fiege-2009} to circumvent it. 

\subsection{Equilibration and data aquisition}

For packing fractions up to $\eta=0.8$ we used the same data sets from \cite{gholami2011slow} as initial conditions. For even denser systems, we used a compactified system acquired as described in section \ref{sec:comp}.

The data shown in section \ref{sec:gamma} were obtained by first equilibrating 100 different configurations and then switching on the force on the intruder. The final velocity of the intruder was measured after stationarity was attained. The time to reach a stationary state depends on the packing fraction, the restitution coefficient and the applied force on the intruder. A typical trajectory and its corresponding fluctuating velocity for the largest packing fraction $\eta=0.8$ and force $F=100000$ is shown the inset of Fig. \ref{fig:mobility2}. The intruder moves approximately $3R_0$ in $x$-direction before fluctuating around its stationary average velocity.

\section{Results}
\label{sec:results}

The time rate of change of the total momentum $\mathbf
P=\frac{1}{N}\sum_i m_i\mathbf v_i$ follows from 
Eq. (\ref{eq:com2}):
\begin{equation}
\dot{P}^{\alpha}(t)=-\gamma P^{\alpha}(t)+\frac{F}{N}\delta_{\alpha,x}
\end{equation}
which is solved by
\begin{equation}
P^{\alpha}(t)=P^{\alpha}(0)e^{-\gamma t}+\delta_{\alpha,x}
\frac{F}{\gamma N}(1-e^{-\gamma t}) \label{eq:p_total}
\end{equation}
Here we have used that the collisions as well as the random driving
conserve momentum.
In the frictional case with $\gamma\neq 0$, the total momentum goes to
a constant, $P_x=F/(N\gamma)$, whereas for the frictionless case with
$\gamma= 0$, the momentum grows linearly with time, $P_x=Ft/N$. In the
following we shall mainly discuss the first case, because the
frictional model is closer to experiment. However, the frictionless
case allows us to study the propagation of momentum in an inelastic
fluid \cite{jabeen2010universal,gomez2011shock}.

% The data was obtained by first equilibrating 100 different runs and
% then switching on the force on the intruder.

\subsection{Frictionless state ( $\gamma=0$ )}
\label{sec:comp}

\begin{figure}
  \psfrag{y}{$z$}
  \psfrag{x}{$x$}
  \includegraphics[width=0.5\textwidth]{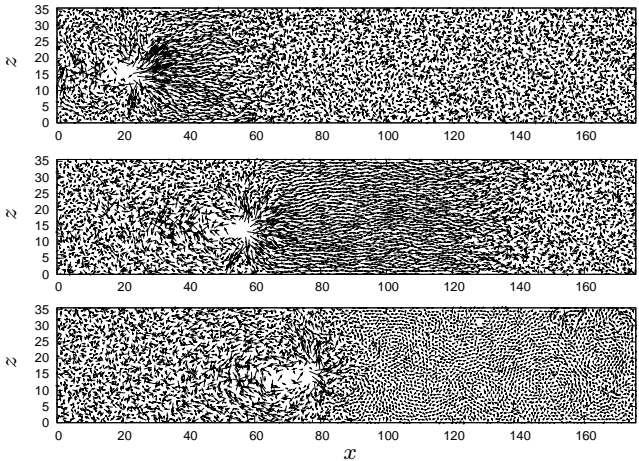}
  \caption{Momentum transport with $\gamma=0$ and hard walls. A vector
    denoting the particle's velocity is assigned to each particle's
    position. ($T = 0.1$, $\epsilon = 0.9$, $F = 500$, $\eta=0.8$, $t=2.6$, $6.3$, $8.5$)}
  \label{fig:mom}
\end{figure}

To that end, we consider a system with aspect ratio $A=5:1$, $N=5000$ and fixed
walls, which reflect the particles elastically. 
Momentum is conserved on average except for the pulling force which
constantly feeds (small) momenta into the system, which are propagated
by collisions away from the source into all of the available area. 
Below the jamming transition momentum transport is given by ballistic
motion of particles as well as by collisions. If the intruder starts
at the left hand side of the system and is pulled by the external
force, then only particles in the neighbourhood of the intruder will
feel the local momentum input for short times. The momentum
given to the intruder is distributed among the particles in front
(i.e. in the direction of the pulling force) of the intruder. Again,
these particles distribute their momenta by collisions as well as by
ballistic transport. A front of particles carrying the momentum fed
into the system by the intruder is formed (see Fig. \ref{fig:mom}
top), propagates through the system (see Fig. \ref{fig:mom} center)
and ultimately collides with the hard wall (see figure \ref{fig:mom}
bottom). At this instant the momentum is reflected by the wall and
many particles end up in a highly compactified state with a packing
fraction $\eta \approx 0.839$. 

\begin{figure}
  \psfrag{t20}{$t=1.3125$}
  \psfrag{t60}{$t=3.8125$}
  \psfrag{t100}{$t=6.3125$}
  \psfrag{px}{$P^x$}
  \psfrag{x}{$x$}
  \includegraphics[width=0.5\textwidth]{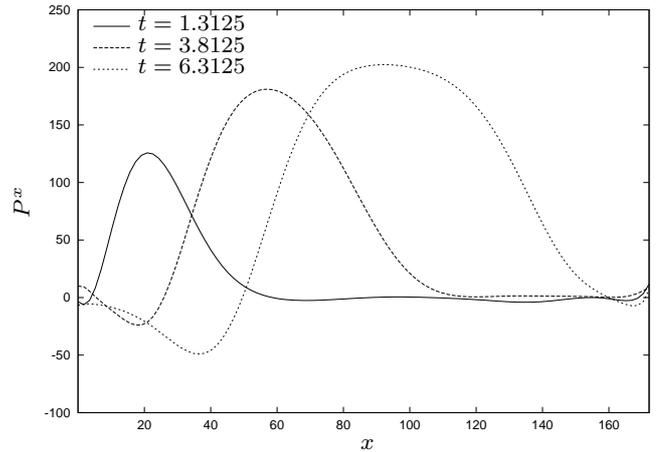}
  \caption{Propagation of momentum wave; total momentum for 3
    different times, in dimensionless units; (parameters as in
    Fig. \ref{fig:mom}).} 
  \label{fig:propagation}
\end{figure}

To analyze the momentum wave quantitatively, we plot $P_x$ averaged
over $z$ in Fig. \ref{fig:propagation}. The propagation front is well
defined, its center of mass moves with velocity $V_{\mathrm{cm}}=16.51$ and it
broadens with time such that it's width increases linearly with time,
$\Delta=V_{\mathrm{br}}t$ (with $V_{\mathrm{br}}=10.89$). Both velocities increase with pulling force, while their ratio is approximately constant. Hence the observed propagation front cannot be identified with linear sound but is presumably a shockwave in agreement with the observations of Ref. \cite{gomez2011shock}.


\subsection{Force-velocity relation ( $\gamma\neq 0$ )}
\label{sec:gamma}

In the following, we consider a system with $N=20000$ grains, subject to
friction ($\gamma =1$) with hard walls in the $z$-direction
and periodic boundary conditions in the $x$-direction, allowing for a
stationary current.
We analyze the steady state motion of the intruder for small and large
forces as a function of packing fraction for $T=1$ and subsequently
compare these results to the same system with $T=0.04$. 
% A typical
% trajectory, together with the fluctuating velocity of the intruder is shown in the inset of
% Fig. \ref{fig:mobility2}.

\subsubsection{Linear Regime: Mobility of the intruder}
\begin{figure}[h]
%   \psfrag{mu: eps=0.9}[c]{$\mu(\epsilon=0.9)$}
%   \psfrag{mu: eps=0.7}[c]{$\mu(\epsilon=0.7)$}
%   \psfrag{mu}{$\mu$}
%   \psfrag{t}{$t$}
%   \psfrag{x}{$x$}
%   \psfrag{v}{$v$}
%   \psfrag{phi}{$\eta$}
  \includegraphics[width=0.5\textwidth]{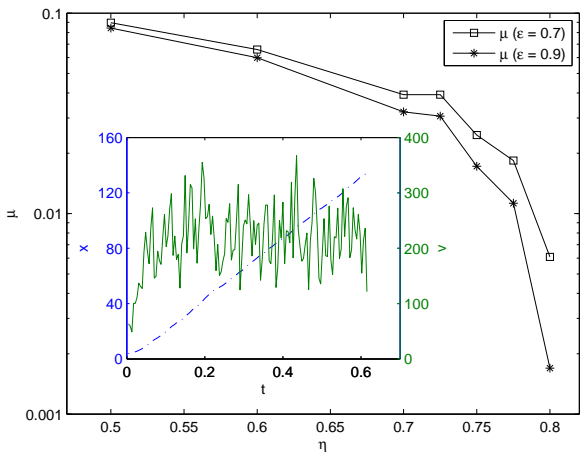}
\caption{Mobility as a function of volume fraction for $\epsilon=0.9$
  (stars) and $\epsilon=0.7$ (squares). Inset: Typical time dependence of the $x$ position and $x$ velocity of the intruder for $\epsilon =0.9$, $F=100000$ and $\eta=0.8$.}
 \label{fig:mobility2}
\end{figure}

For a small pulling force we expect a linear relation between the
velocity of the intruder and the force:
\begin{equation}
v_I=\mu F,
\end{equation}
defining the mobility $\mu$ of the intruder.
\begin{figure}[h]
  \psfrag{F}{$F$}
  \psfrag{v}{$v$}
  \psfrag{030}{$\eta = 0.3$}
  \psfrag{050}{$\eta = 0.5$}
  \psfrag{060}{$\eta = 0.6$}
  \psfrag{070}{$\eta = 0.7$}
  \psfrag{0725}{$\eta = 0.725$}
  \psfrag{075}{$\eta = 0.75$}
  \psfrag{0775}{$\eta = 0.775$}
  \psfrag{080}{$\eta = 0.8$}

  \includegraphics[width=0.5\textwidth]{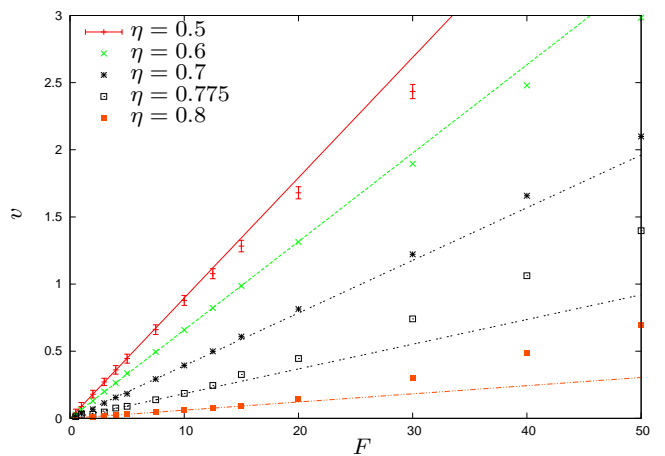}
   \caption{Velocity versus pulling force for packing fractions
    $0.3\leq\eta\leq0.8$, parameters chosen: $\gamma
    =1,\epsilon=0.7$ and aspect ratio $A=1:2$.}
  \label{fig:mobility}
\end{figure}
In Fig. (\ref{fig:mobility}) we plot the velocity of the intruder versus
force and indeed do observe a linear regime for small pulling force.
From the slope we extract the mobility which is shown in
Fig. (\ref{fig:mobility2}).

The breakdown of the mobility when the glass transition is approached is clearly visible for both investigated $\epsilon$. The more inelastic system with $\epsilon=0.7$ shows increased mobilities as compared to $\epsilon =0.9$. Since the particles are more ``sticky'', they tend to stay closer after a collision than in the elastic limit, i.e., stronger density fluctuations are inherent to systems with stronger inelasticity. This enlarges the accessible space for the intruder compared to the system with $\epsilon =0.9$, increasing its average velocity. 
Moreover, the momentum of the intruder in the direction of the pulling force is reduced due to collisions with other particles in front. In the center of mass frame, the intruder is reflected backwards. For the more inelastic system this kind of backscattering is less effective (see Eq. \ref{eq:coll1}), so that the velocity in the direction of the force and hence the mobility is larger for the more inelastic system.

%These results confirm the data from a previous study \cite{gholami-2011} where the diffusion coefficients in a nonfrictional system were studied. 

\subsubsection{Nonlinear regime}

\begin{figure}[h]
\psfrag{x}{$\propto F^{\beta}$}
\psfrag{F}[cc]{$F/(v_{\mathrm{dr}}^2 \cdot \eta^{1/2} )$}
\psfrag{v}[l]{$v/v_{\mathrm{dr}}$}
\psfrag{030}[r]{$\eta = 0.3$}
\psfrag{050}[r]{$\eta = 0.5$}
\psfrag{060}[r]{$\eta = 0.6$}
\psfrag{070}[r]{$\eta = 0.7$}
\psfrag{0725}[r]{$\eta = 0.725$}
\psfrag{075}[r]{$\eta = 0.75$}
\psfrag{0775}[r]{$\eta = 0.775$}
\psfrag{080}[r]{$\eta = 0.8$} 
\psfrag{xb}[r]{$F^{\beta}$} 
\includegraphics[width=0.5\textwidth]{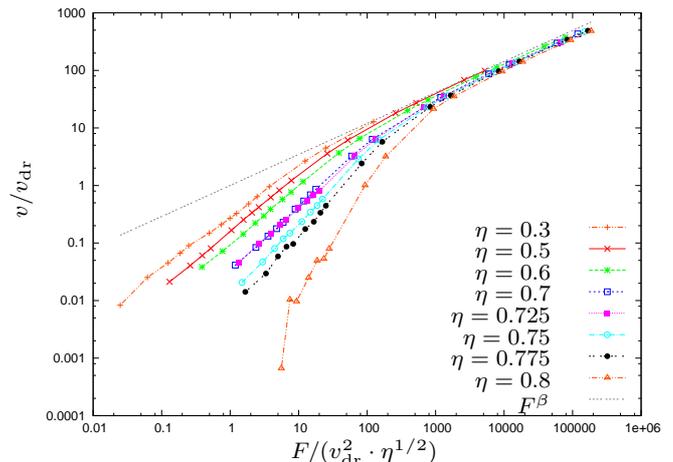}
 \caption{Velocity-force relation for all pulling forces, $\epsilon =0.9$.}
  \label{fig:mom2}
\end{figure}

As the pulling force is increased, deviations from linear behaviour
are expected and observed. To investigate these systematically we have
applied forces in the range $1 \leq F\leq 10^5$ and show our data in a
scaling plot in Fig. (\ref{fig:mom2}). Scaling velocities by
$v_{\mathrm{dr}}=p_{\mathrm{dr}}/m_{\mathrm{eff}}$ and forces by $v_{\mathrm{dr}}^2\cdot \eta^{1/2}$
collapses the data for large forces. The velocity of the
intruder scales algebraically with the pulling force, according to
$v_I\propto F^{\beta}$, $\beta=0.55$.

The crossover between the linear and the algebraic regime is accompanied by range of forces in which the intruder velocity increases superlinearly with the applied force. This is observed at high packing fractions only, see Fig. \ref{fig:mobility}. We conjecture that this \emph{shear thinning} is due to the formation of vortices which can only be formed at high packing fractions, where momentum is conveyed almost instantaneously as the disks are very close to each other. For low packing fractions, the distance of neighbouring particles is too large for the formation of vortices because the damping $\gamma$ dissipates most of the momentum.

% \begin{figure}[h]
% \includegraphics[width=0.5\textwidth]{vertex.eps}
% % \includegraphics[width=0.5\textwidth]{neue_bilder/eps07collabs12.eps}
%  \caption{Vortex formation for a system with $\epsilon =0.7$, $\eta=0.775$, $F=50$, $T=1$. The intruder is the green disk in the middle, the remaining disks are color-labelled according to their kinetic energy, where clear red corresponds to $E_{\mathrm{kin}} \geq 5$, white to $E_{\mathrm{kin}} =0$, intermediate colors accordingly; arrows depict the velocity of each particle. A time dependent vortex forms below the intruder, carrying momentum from its front to its back. The time difference between two figures is $0.1129$, in which roughly 1 collision per particle takes place.}
%   \label{fig:vortexes}
% \end{figure}

\subsubsection{Temperature dependence}

The crossover from linear to nonlinear response depends on the thermal
velocity, $v_{\mathrm{th}} \simeq \sqrt{2T/m_0}\approx 0.7$. For small
packing fractions, this crossover actually occurs at $v_{\mathrm{th}}$
as can be seen e.g. in Fig. \ref{fig:mobility} for the smallest
packing fractions. Decreasing the temperature is expected to shrink
the linear regime also for higher densities. Hence we try to explore
the force velocity relation in the same system but with smaller
thermal velocity. Here we choose $T=0.04$. For low pulling forces, we
expect the mobility to be larger than in the case $T=1$ since the
decreased thermal motion does not disturb the intruder travelling
through the almost resting surrounding disks. For high forces, we
expect the intruder velocity to not depend on the temperature, since
in this case, the intruder's velocity is at least one order of magnitude
higher than the thermal velocity (see Fig. \ref{fig:mom2}).

\begin{figure}[h]
 \includegraphics[width=0.5\textwidth]{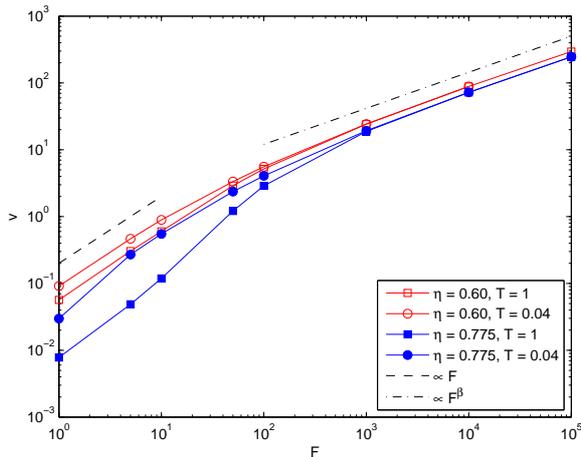}
 \caption{Velocity-force relation for strong pulling forces at different temperatures and $\epsilon=0.9$.}
  \label{fig:T_low}
\end{figure}

These expectations are indeed born out by the data. We plot the force
velocity relation for packing fractions $\eta = 0.6$ and $0.775$ for
both temperatures in Fig. \ref{fig:T_low}. For high pulling forces $F
\geq 10^3$, the intruder velocities for low and high temperature
collapse for both packing fractions as expected. For $\eta=0.6$ and
$T=0.04$, the crossover from the linear regime to the nonlinear regime
is shifted to smaller forces. For $\eta=0.775$ and $T=0.04$, the linear
regime is not even detectable. The crossover is shifted by roughly two
decades. To explore the linear regime would require forces
as small as $F=10^{-2}$, at which the average intruder velocity would
be too small to be detectable.

\section{Conclusion and Outlook}
\label{sec:concl}

We have generalised an event-driven algorithm to include friction. As
a first application, we have studied the dynamics of an intruder
pulled by an external force. In contrast to previous simulations
\cite{reichhardt2010}, we consider a fluidized granular medium, which
is expected to undergo a glass transition at a packing fraction below
random close packing \cite{kranz2010}. We do indeed find a dramatic
decrease of the mobility around $\eta=0.8$, in agreement with previous
simulations without pulling force and consistent with a glass
transition. For large pulling force, the data can be collapsed by
scaling, following a power law dependence $v_I\propto F^{\beta}$ with
$\beta=0.55$.

In the frictionless case the pulling force generates a momentum wave
propagating through the sample and thereby compactifying it. We plan
to investigate momentum transport in more detail in the
future. Furthermore the generalised event driven algorithm will be
useful more generally in the context of frictional granular matter
fluidized by air or water flow. Work along these lines is in progress.

%\input{concl.tex}
%\begin{figure}
%  \includegraphics{example.eps}
%\end{figure}
%\begin{figure*}
%  \includegraphics[width=0.75\textwidth]{example.eps}
%\end{figure*}
%\begin{table}
%\begin{tabular}{lll}
%\hline\noalign{\smallskip}
%\end{tabular}
%\end{table}

\begin{acknowledgements}
We thank T. Aspelmeier, I. Gholami, C. Heussinger, T. Kranz and S. Ulrich for many useful discussions. We furthermore acknowledge support from the DFG by FOR 1394.
\end{acknowledgements}

\bibliographystyle{spphys}       % APS-like style for physics
\bibliography{lit.bib}   % name your BibTeX data base

%\begin{thebibliography}{}
% etc
%\end{thebibliography}

\end{document}